\input epsf
\documentstyle[12pt,cites,doublespace]{article}
\newfont{\kleiner}{eurm10}
\begin{document}
\vspace*{0.8cm}
{\noindent  
 ELECTRON CORRELATIONS IN THE HIGH $\rm T_c$-COMPOUNDS}\\

{\noindent Gernot STOLLHOFF}\\
Max-Planck-Institut f\"ur 
Festk\"orperforschung, D-70569 Stuttgart,
Germany \\


\begin{singlespace} 
\noindent
Ab-initio correlation results for an idealized high $T_c$-compound
are compared to density functional (DF) calculations for the same system.
It is shown that and why the DF-charge distribution is wrong. The
largest deficiency arises for the $Cu$-$d_{x^2-y^2}$-occupation,
arising from strong atomic correlations but mostly from anomalous neighbor 
$Cu$-spin correlations. Both features are beyound the range of the
homogeneous electron gas approximation underlying the DF-schemes.
The ab-initio results also exclude a description of the real
system in a Mott-Hubbard scenario, that is mostly chosen in theory.
Conditions for models are derived that are able to describe the
high-$T_c$-compounds.\\
\end{singlespace}

{\noindent 1. INTRODUCTION}

In the past, two approaches have dominated the attempts to
understand the electronic structure of the high-$T_C$ compounds.
The first is by ab-initio DF-calculations within the local density
approximation (LDA), and the second is by very particular models.

The LDA calculations suffer from the underlying homogeneous electron gas 
approximation for exchange and correlations. They are able to represent
properties like equilibrium lattice constants, phonons, or the
Fermi surface, but are unable to reproduce the magnetic phase at half 
filling, and have deficiencies for the band masses. The obtained
electron lattice coupling is too small for the explanation
of superconductivity. 

The particular models are mostly restricted to  
the $Cu$-$O$-planes. The description of the electrons in
these planes is usually further restricted to a single band Hubbard model
in the strong correlation limit or to a t-J model. Such a model can explain
the magnetism of the so called half-filled planes, and deliberate
Fermi surfaces can be generated by particular parametrizations. A
connection to the microscopic reality however, or even the unequivocal
construction of a plausible new mechanism of superconductivity, could
not yet be obtained.

We had been able to pursue a different approach that 
neither suffers from the uncontrollable shortcomings of
the LDA, nor from an
equally uncontrollable ad hoc choice of a particular
model. With the help of the Local Ansatz (LA), we were able 
to perform an ab-initio correlation
calculation for an idealized high-$T_c$ compound$^1$.

It is generally agreed that a basic understanding for
the high $T_c$-compounds can be gained from
the treatment of a simplified idealized compound, namely a
single charged $Cu$-$O$-plane. Actually most models make
use of this approximation. 
Our calculation was restricted to such a system, or rather to its
so called infinite layer generalization into three dimensions.

In the following, we will compare the LA-results to similar
LDA-calculations, and will unequivocally determine
the deficiencies of the LDA, and their causes. We will also
depict a few important correlation results and give a critical
valuation of the most often used models.\\

\noindent 2. COMPUTATION METHOD

The LA allows the
ab-initio computation of the correlated
ground state of a solid. Starting point for the correlation treatment
is a self consistent field (SCF) or Hartree-Fock calculation for
this solid, resulting in the uncorrelated single-particle
ground state $\Psi_{SCF}$. This SCF-calculation was performed
by the program Crystal$^2$. The latter program uses atom centered 
Gauss type orbitals (GTO) as a basis, as does the LA.

Within the LA,
the following variational ansatz is made for the correlated ground
state:
\begin{eqnarray}
| \Psi_{\rm corr} \rangle & = & e^{-S} | \Psi_{\rm SCF} \rangle 
\label{eq:lanst0} \\
      S                   & = &  \sum_{\nu} \eta_{\nu} O_{\nu}   
\label{eq:lanst01} \\
O_{\nu}                   & = &  \left\{ \matrix{
                                      n_{i \uparrow} n_{i \downarrow} \cr
                                      n_i n_j                         \cr
                                      \vec{s}_i \cdot \vec{s}_j     \cr
    \{n_{i \uparrow} (a_{i \downarrow}^{\dagger}a_{j\downarrow}-
a_{j \downarrow}^{\dagger}a_{i\downarrow})\} +
\{ \uparrow \leftrightarrow \downarrow \} \cr
n_i } \label{eq:lanst}
                                                     \right. \ \ .
\end{eqnarray}
The $\eta$'s serve as variational
 parameters. The $n_{i \sigma}$ and $\vec{s}_i$ are density and spin operators for an
electron in the local state $a_{i \uparrow}^{\dagger}$, represented by
the orbital
\begin{equation}
g_i(\vec{r}) = \sum_j \gamma_{i j} f_j(\vec{r})       \label{eq:local}
\end{equation}
where the $f_i(\vec{r})$ are the (GTO like) basis orbitals. 
The operators have an
obvious meaning. The first operator $n_{i \uparrow} n_{i \downarrow}$,
for example, when applied to $ | \Psi_{\rm SCF} \rangle $, projects out all
configurations with two electrons in orbital $ g_i(\vec{r})$. In connection
with the variational parameter $ \eta_{\nu}$, it
partially suppresses those configurations. Similarly, the operators $n_i n_j$
describe density correlations between electrons in local orbitals
$g_i(\vec{r})$ and $ g_j(\vec{r})$.
The operators $ \vec{s}_i \cdot \vec{s}_j $ generate
spin correlations. The fourth kind of operators is of the form of
$[O_{\nu},H_0]_-$, where $H_0$ represents the single-particle Hamiltonian.
These operators refine the
ansatz with respect to the band energy of the electrons involved. 
The original operators of eq. 
\ref{eq:lanst} are next modified by subtracting the
 contracted contributions in each of them.
The corrected operators when applied to $ | \Psi_{\rm SCF} \rangle$
contain only two-particle excitations, and the corrected
last kind of operators in eq. \ref{eq:lanst} covers local single 
particle excitations, i.e. it allows for changes in occupations.

   The variational parameters $\eta_{\nu}$ are chosen to optimize
the energy
\begin{eqnarray}
E_G & = & \frac{\langle\Psi_{\rm corr} | H | \Psi_{\rm corr} \rangle}
               {\langle\Psi_{\rm corr} | \Psi_{\rm corr} \rangle} \\
    & = & \langle\Psi_{\rm corr} | H | \Psi_{\rm corr} \rangle_{c} \ \ .
\end{eqnarray}
In the last equation,
 the subscript ${}_{c}$ indicates that only connected
diagram contributions are summed up.
This expression cannot be evaluated exactly.
The standard approximation is an expansion in powers of $\eta$, up to second
order,
\begin{eqnarray}
E_G      & = & 
       E_{\rm SCF} + E_{\rm corr}                 \\
E_{\rm corr} & = &  - \sum_{\nu} \eta_{\nu} \langle 
O^{\dagger}_{\nu} H\rangle \\
     0 & =  &  - \sum_{\nu} \eta_{\nu} \langle O^{\dagger}_{\nu} H\rangle
     + \sum_{\nu,\mu} \eta_{\nu} \eta_{\mu} \langle O^{\dagger}_{\nu}
HO_{\mu}\rangle_c
     \ \ .               \label{eq:expan}
\end{eqnarray}
Here, $\langle A \rangle$ means the expectation value of the operator
$A$ within
$| \Psi_{\rm SCF}\rangle$. 
This is a weak correlation approximation that blows up when
correlations turn too strong. For the high-$T_c$'s no such problems
arose, indicating that the Mott-Hubbard limit does not apply.

The local orbitals in eq.~\ref{eq:local}
are connected to a single atom only and are built from
its basis orbitals. Standard Quantum Chemistry
(QC) methods also start from a SCF-calculation and add correlations 
with the help of one-and two-particle operators. However, the QC-operators
are constructed from orthogonal sets of occupied and/or unoccupied
orbitals, a construction that fails for metals. The restriction to
local orbitals and the particular construction of correlation operators
from these local orbitals is the essential approximation of the LA 
in comparison tho QC. It allowes even to treat metals with QC accuracy.

For the high-$T_c$ application, only operators built from
atomic orbitals were used.  The atomic orbitals are 
unequivocally determined from the SCF-ground state
by the condition that
they are built from basis orbitals on the respective atoms only and
that they cover a maximal fraction of the full occupied space.
The resulting orbitals are next L\"owdin-orthogonalized to each other.
More  localized subatomic orbitals were not used in this applicaltion.

The calculated system is formally be described as $SrCuO_2$.
It is half-filled and has an antiferromagnetic
ground state. However, in all calculations, not this
antiferromagnetic ground state but a metastable non magnetic state
was treated since we are not interested in the magnetic
order but in the electronic properties of the doped high-$T_c$ compounds.
This approach is necessary because the Crystal program requires
unit cells with integer filling, but we wanted to avoid large unit
cells. We took care that
none of the computed results was influenced by the proximity to the
magnetic instability.

Being interested on in-plane features, we choose a good basis for
the in plane $Cu$- and $O$- atoms, but added no valence basis orbitals 
for the $Sr$-atoms. This leads to a charging of the planes
with 2 electrons per $Cu$-atom, and renders the interplane coupling
negligible.
For more details of the basis choice, of the structure of the idealized
system, and of the correlation calculations, we refer to ref. 1).\\

\noindent 3. ATOMIC CHARGE DISTRIBUTION

Being unequivocally defined, the atomic orbitals, generated
for correlation purposes, 
are also a perfect basis for a detailed charge
analysis. This representation avoids all non orthogonality 
problems of a standard
Mulliken population analysis.

The partial charge distributions 
$n_i(\Psi)=\langle\Psi| \sum_{\sigma}n_{i\sigma}|\Psi\rangle$
are presented in table \ref{tab32} for different states $\Psi$.
The first row contains the values for $\Psi=\Psi_{SCF}$.
The sum over the partial occupations approaches the
number of valence electrons within 0.02. 
This proximity rectifies the condensation into
atomic orbitals and the specific method of their computation.

With the addition of correlations, a relatively large charge transfer 
occurs. Ultimately, it is a charge transfer mostly from the
$Cu 3d_{x^2-y^2}$-orbitals into the $O 2p$-orbitals. However, for
its understanding it is necessary to progress stepwise.
A first step is the addition of atomic correlations which
lead to a large correlation energy gain.
The dominant charge transfer due to the atomic correlations
is from the $Cu 3d_{x^2-y^2}$-orbitals
to the $Cu 4s,4p$-orbitals, followed by a secondary 
redistribution
from the $Cu 4s,4p$-orbitals to the $O 2p$-orbitals.
Over all, 0.18 electrons are removed from the 
$Cu 3d_{x^2-y^2}$-orbitals, and put into the $Cu 4s,p$-shell (0.13) and
the $O 2s,p$-shells (2x0.03).
More than 80 percent of this charge transfer arise from the inclusion
of the operators
$n_{i\uparrow}n_{i\downarrow}$ for the $Cu 3d_{x^2-y^2}$-orbitals, the
remaining part stems from the same operators for the $4s,p$-orbitals.
For an explanation, we refer to ref 1).

When neighbor correlations are included, then an additional charge
transfer of the same magnitude as the one due to on-site correlations
occurs.
It is dominantly from the
$Cu 3d_{x^2-y^2}$-orbitals to the $O 2p_b$-orbitals, and is due to
a particular spin correlation between neighbor $Cu$-sites that will
be discussed later. The longer range contributions that were covered
by the present computations lead to a further but small transfer
of the same kind. For an error analysis, we refer to ref. 1) again.

\begin{table}
\begin{center}
{   \begin{tabular}{|l|r|rrr|lc|}
 \hline                                                                        
Orbital& HF&on site corr&nn corr&full corr&LDA&\quad\\
\hline
$Cu 3d_{x^2-y^2}$ & 1.51& 1.33& 1.17& 1.15&1.53&\\
$Cu 3d_{z^2}$ & 1.95& 1.94& 1.94& 1.94&1.90&\\
$Cu 3d_{xy},3d_{xz},3d_{yz}$ & 2.00& 2.00& 2.00& 2.00&1.99&\\
\hline
$Cu 4s$ & 0.50& 0.55& 0.57& 0.58& 0.64 &\\
$Cu 4p_{pl}$ & 0.30& 0.33& 0.34& 0.34&0.31&\\
$Cu 4p_{\perp}$ & 0.09& 0.11& 0.11& 0.11 
&0.18&\\
\hline
$O 2s$ & 1.82& 1.82& 1.81& 1.81&1.80&\\
$O 2p_b$ & 1.42& 1.48& 1.57& 1.58&1.39&\\
$O 2p_{orth}$ & 1.97& 1.96& 1.96& 1.96&1.94&\\
$O 2p_{\perp}$ & 1.95& 1.93& 1.92& 1.91&1.92&\\
\hline
\end{tabular}}                                                                 
\end{center}\protect
\caption{
Charge distributions for the SCF ground state and with correlations
added, in comparison to LDA results. The subindices of the $p$-orbitals
have the following meaning: $\perp$ perpendicular to plane,$pl$ in plane,
$b$ in bond direction, $orth$ in plane perp. to bond} 
\label{tab32}
\end{table}

The latest version of the program Crystal$^2$ also allowes to perform
LDA-calculations within the same basis set as used for the 
SCF-calculation. The resulting LDA charge distribution is
analysed in the same way as done before for the SCF-case. 
Most alternative
LDA-schemes have no atom centered basis. Thus, 
their ground state results can not be fully decomposed
into atomic occupations. Only
partial results were so far available, mostly from
tight binding fits to the energy bands. Here,
the first full LDA charge analysis is given. 

The LDA-result is included in table \ref{tab32}. 
The occupations of the $3d$-orbitals
are in very good agreement with
earlier LDA results (for citations, see Ref 1)).
There is a close agreement with the SCF-result. Only 
orbitals almost filled in LDA are even
more filled in SCF-approximation, resulting from the well known
band spreading of the SCF. Correlations partially undo this.
The proximity to SCF indicates that none of the 
correlation corrections are contained in the LDA result. 
It is plausible that the anomalous
neighbor correlation effects can not be covered by the homogeneous electron
gas approximation, but it was somewhat astonishing that the
atomic correlation effects are also completely disregarded in LDA.

These occupation results
demonstrate why LDA must be very deficient for the
high $T_c$-compounds.
A consequence for the band structure can be easily derived.
When keeping the LDA hopping terms in the single-particle Hamiltonian
but shifting the crystal field terms so that the resulting single-particle
ground state has the correct charge distribution, then 
the Fermi surface stays the same, but the conduction
band shrinks by half, bringing it into much closer agreement to 
experiment. More details will be given elsewhere$^3$.\\

\noindent 4. SPIN CORRELATIONS AWAY FROM HALF FILLING

All LA results discussed so far were
connected to LDA results. But in contrast to the LDA, the LA
obtains also full information
for the correlated ground state, and in particular
correlation functions. In the following, we will deal
with the surprising
anomalous neighbor $Cu$-spin
correlations. These are connected to
a charge transfer from the two $Cu$-atoms into the inbetween $O$-atom
because the latter enhances the antiferromagnetic coupling and further 
reduces the charge fluctuations on all three
involved atoms. This is a correlation feature of the delocalized
electron system. It is not at all connected to a Mott-Hubbard
transition which can be ruled out by our results.
These spin correlations
mutually enhance each other. Even without
any long range magnetic order in the metastable metallic state, they are so
big that the repulsive effect of the on-site $Cu$-correlation hole
is overcompensated, and that electrons with differing spins have a higher
probability to be close to each other (i.e. up to  neighbor sites)
than without correlations. 

This finding has a similarity
to the so called renomalized valence bond (RVB) 
picture that had been proposed as a mechanism for 
superconductivity$^4$, and that also 
leads to a strong neighbor spin correlation . 
While the RVB approach is only valid close to the Mott-Hubbard
transition, our results
arise already for weak correlations and are also present at large dopings. 

These short range magnetic correlation
features are next compared to experiment. 
This can only be done outside the magnetically
ordered phase. While the ab-initio calculations themselves can not be extended
to fractional charging due to the Crystal program limitations,
a valuable alternative is to condense the half-filled case informations 
to a model, 
and to extend this model to differing
fillings. Such a model needs to describe the conduction band well,
and also needs to explicitly include the electronic interactions.
We had chosen the smallest possible model that contains the 
$Cu$-$3d_{x^2-y^2}$ orbitals, the $O$-$2p_b$ orbitals, and also
the $Cu$-$4s$ orbitals. The latter orbitals were needed
to obtain a correct occupation of the other two orbitals, necessary
for the short range spin correlations.
A three band model (without the $4s$-orbitals) turned
out to be very deficient. Even our model choice caused shortcomings
because we had to move charge from the $4s$-orbitals into the
$2p$-orbitals in order to preserve the correct Fermi surface. This
might be overcome in the future by extending the model by
$4p$-orbitals. 

The hopping terms for this model were taken
from an LDA-band structure fit found in the literature$^5$, and the
crystal field terms were obtained from the required occupations.
As interaction, Hubbard interaction terms $U_i$ were included for
the individual orbitals. Their value was set so that the
atomic correlation functions obtained by the LA for the model
were the sames as the ab-initio results$^1$.
This model describes the atomic correlations correctly but gives,
probably for the deviations in the occupations, too small neighbor
spin correlations. We therefore performed a parallel calculation
with a $U_{3d}$ that was enhanced by 20 percent.

With this model we calculated
the spin correlation function
\begin{equation}
S(\vec{Q})
=\sum_{i,j,\vec{G} }
\langle\Psi_{corr}|\vec{s}_i(0)\vec{s}_j(\vec{G})|\Psi_{corr}\rangle 
e^{i\vec{Q}(\vec{r}_i-\vec{r}_j-\vec{G})} \ \ \ .
\label{eqs3}
\end{equation}
of the model with 15 percent doping. Here, $i,j$ runs over the atomic 
orbitals 
on sites $r_i,r_j$ 
in the unit cell, and $G$ describes the lattice summation.
This function can be compared to the corresponding, experimentally
measured spin
correlation function$^7$ $S(\vec{Q})$ for the 
metallic compound $La_{0.85}Sr_{0.15}Cu_2O_4$.
The latter is no equal time correlation function but only energy integrated
up to 0.15eV.

\begin{figure}[hbtp]
\epsfysize=65mm
\epsfxsize=75mm
\centerline{\epsffile{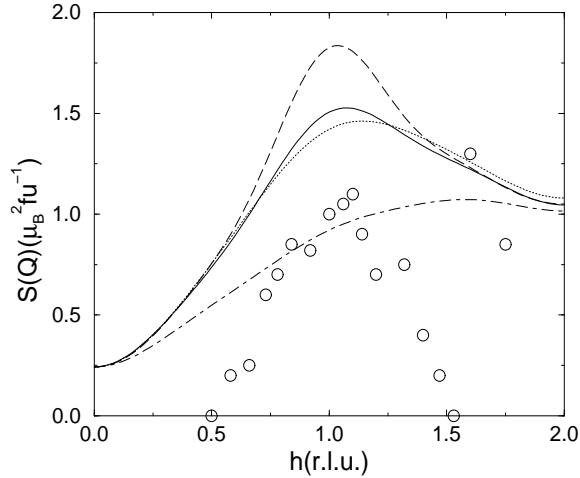}}
\caption{\protect Equal time spin correlation function 
S(Q) for $\vec{Q}=(h,h,0)$
in comparison to experiment$^6$ (empty circles).
Given are the results of the HF-ground state (broken-dotted curve),
the 5 atom cluster result (dotted line) and the 9 atom cluster result
(continuous line) for  $U_{3d}$=6.3eV, and the 9 atom 
cluster result for
$U_{3d}$=7.8eV.}
\label{figs1}
\end{figure} 
\vspace{0.5cm}
Fig. \ref{figs1} contains the experimental and theoretical
results for 
the diagonal (1,1) axis. 
The zone boundary is at h=1, the intensity is
given per formular unit which here is equivalent to a unit cell or to a
single $Cu$ atom. 

The lowest curve represents the result for the single-particle ground
state. It represents the exchange holes. As the finite value at $h=0$
indicates, the summation in eq. \ref{eqs3} was not brought to convergency.
The maximal deviation occurs for $h=0$ were the contributions from all 
missing terms add up. Due to dephasing, the correction is very much
smaller for finite $h$. 

Next, short range correlations as they are deduced from a single coherent
5 $Cu$ cluster calculation are included (dotted curve). Here, the nearest
neighbor $Cu-Cu$ correlations come into play and cause a peak at
the zone boundary ($h=1$).
When extending the correlation treatment to a 9 Cu
cluster, the peak narrows somewhat and enhances a little (continuous
curve). Finally, also the corresponding values with enlarged $U$ (7.8eV
instead of 6.3eV) are given (broken curve).

As expected, the
theoretical equal time correlation function is always larger
than the experimental correlation function whose energy integration extends
only to 0.15 eV. One would expect that a sizable part of the correlation 
function, namely the one already obtained by $\Psi_{SCF}$ can 
definitely not be
seen by the slow neutrons. Consequently, all remaining
contributions need to be  seen in the experiment. This might 
occur because the electrons partially localize due to a charge density
wave formation in the case of the treated compound.

It should be noted that neither Hubbard model nor t-J-model
results are able to explain such a strong inelastic magnetic
scattering at 15 percent doping. Also a 3 band model would 
definitively fail if reasonable Hubbard interaction terms were chosen.
Even the 4 band model used here displays deficiencies.

Our results indicate that the origin of the inelastic magnetic
scattering might well be completely disconnected 
from the magnetic order at half
filling, and might be explained by the anomalous short range correlations
found over a large range of doping. At least the theoretical Q-dependence
matches the experiment very well. \\

\noindent REFERENCES

\begin{singlespace} \noindent
1) G. STOLLHOFF, Phys. Rev. {\bf B 58} (1998) 9826.\\
2) V.R. SAUNDERS et al., Crystal98 User's Manual, University of 
Torino (1998).\\
3) G. STOLLHOFF, unpublished. \\
4) P.W. ANDERSON, Science {\bf 235} (1987) 1196.\\
5) O. K. ANDERSEN et al., J. Phys, Chem. Solids {\bf 56} (1995) 1573.\\
6) M. HAYDEN et al., Phys. Rev. Lett. {\bf 76} (1996) 1344.\\
\end{singlespace}
\end{document}